# Adapting SAM for CDF

D. Bonham, G. Garzoglio, R. Herber, J. Kowalkowski, D. Litvintsev, L. Lueking, M. Paterno, D. Petravick, L. Piccoli, R. Pordes, N. Stanfield, I. Terekhov, J. Trumbo, J. Tseng, S. Veseli, M. Votava, V. White
*FNAL, Batavia, IL 60510, USA*

T. Huffman, S. Stonjek, K. Waltkins
*Oxford University, Oxford OX1 2JD, UK*

P. Crosby, D. Waters
*University College London, London WC1E 6BT, UK*

R. St.Denis
*Glasgow University, Glasgow G12 8QQ, Scotland, UK*

The CDF and D0 experiments probe the high-energy frontier and as they do so have accumulated hundreds of Terabytes of data on the way to petabytes of data over the next two years. The experiments have made a commitment to use the developing Grid based on the SAM system to handle these data. The D0 SAM has been extended for use in CDF as common patterns of design emerged to meet the similar requirements of these experiments. The process by which the merger was achieved is explained with particular emphasis on lessons learned concerning the database design patterns plus realization of the use cases.

## 1. INTRODUCTION

The CDF and D0 collider experiments have common needs and therefore projects have been established to share a variety of common tools. The adaptation of SAM for CDF data handling requirements was discussed at the end of 2001 in a "prepilot" conceptual study that evolved into a Pilot project (Spring 2002) to make concrete tests by populating the CDF Data File Catalog [1](CDF DFC) into the SAM schema. This demonstrated that the functionality of the CDF DFC was a subset of that which SAM had to offer. Features in SAM concerning distributed computing and cache management were considered important in the near term for CDF which was managing with a single central disk store for its data cache but whose days were numbered.

The CDF DFC is a catalog of files organized in datasets, which contain filesets. The filesets group files in units that had been logical for a specific choice of mass storage and the datasets represent a set of common event types analyzed with certain program versions and calibration sets. This information is encoded in a naming convention that was found to be violated from time to time.

The SAM system is file-based where some process under some condition from some source produces a file. The information about process, versioning, and calibration data used in producing files are explicitly controlled in the SAM database as file meta-data.

The SAM for CDF project was established to determine if the two different philosophies were indeed fundamentally different or could share the same infrastructure.

## 2. GOALS

The project goals for the commissioning of SAM for CDF were the following:

1. Supporting 5 groups to do data analysis. This gave a measure of the size of the grid for the commissioning; there were 3 groups located in the UK, 2 groups in the US and one in Germany.
2. Enabling access to datasets of interest. This grid was to be used to analyze data and produce physics results. It enabled read access to secondary datasets and read/write to higher-level datasets. The target size of the secondary dataset to be considered was 1 TB: it was decided that the STK-Based Silo [2] "STKen" would provide a tape storage capacity of 5 TB for writing data via enstore [3]
3. Production availability of the systems. Key machines were maintained 24x7 and with traditional production support provided by experts.
4. Limited impact on CDF operations. The new sam grid had to have a controllable limited impact on the normal CDF operations. The CDF mass storage, "CDFen", based on DCache [4] and enstore was accessed read-only and with a rate limitation.

## 3. ARCHITECTURE

A CDF central sam station ("fcdfsam" in Fig 1) was installed to pattern the D0 central station. This connected to CDFen for secondary data retrieval (read only) and to STKen for higher-level data read/write. The fcdfsam station served as a router for storing data into enstore (permanent disk in Fig 1), and as cache for files read from STKen requested by the local analysis station (nglas09) and the remote stations. Unlike the D0 central station, D0mino, no central analysis was run on fcdfsam; nglas09 was used instead.

The services needed by SAM included:

1. An Oracle Database server. The machine chosen was the existing CDF offline database server, cdfofprd, a dual processor Sun 4500.
2. A CORBA middleware SAM DBserver, which also held the CORBA naming service, SAM optimizer, web pages and monitoring tools. The machine chosen for this was fndaut, also a SUN 4500.

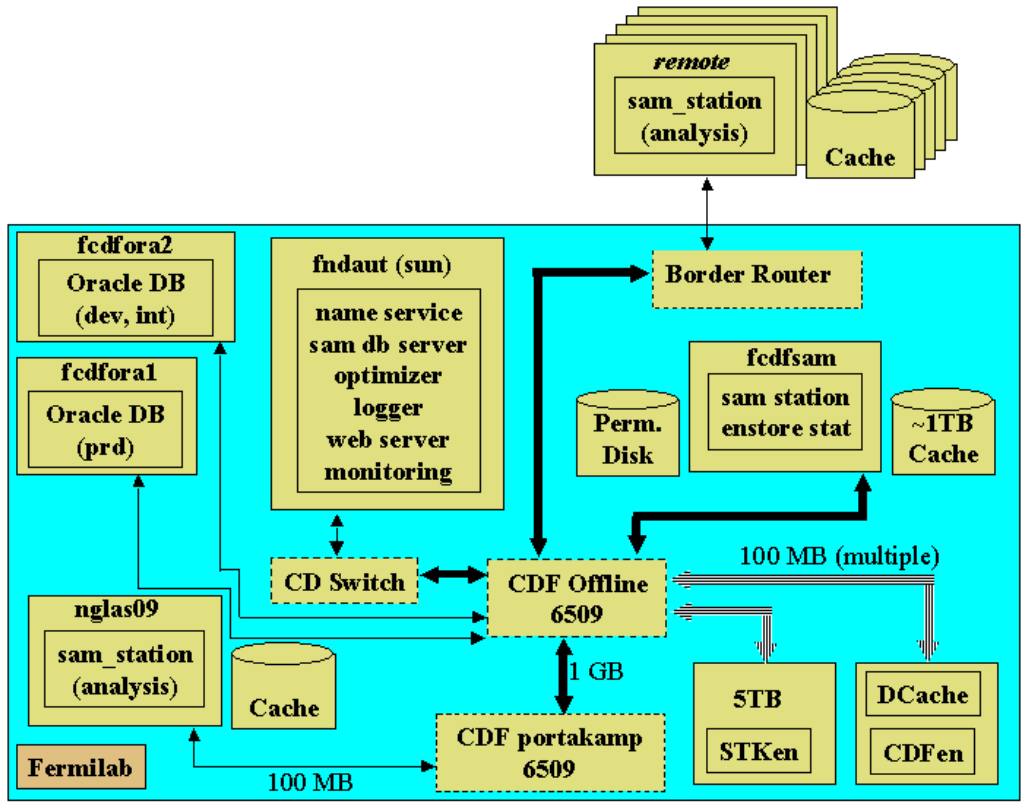

Figure 1: layout of the deployment of the SAM servers for CDF

## 4. DEPLOYMENT

The various items shown in Figure 1 were deployed according to the following plan:

**fcdfsam**: this was located in the computer room on the second floor of the Fermilab Computing Center (FCC2). fcdfsam has 3 1GB ethernet cards. The choice of how to connect fcdfsam depended on the final configuration for the flow of the data between the DCache, the SAM station cache on fcdfsam and the analysis station caches. To accommodate caching on fcdfsam we anticipated a final configuration of 2 dedicated 1 GB cards. In the interim a 1 GB connection was available in the CDF Offline 6509. The disks were configured to allow access to 1.0 Tb of storage across 20 disks. In a matter of weeks after the installation, seven of these disks failed, where the failure was manifested in corruption of the data during transfer. This prompted the storage of checksum quantities in the

SAM database with each file so that all transfers could be checked and retried upon failure. The hardware was replaced with a new machine taken from the CDF CAF [5] having 1.7TB of disk. This disk hosted both the sam cache and the permanent disk needed to buffer file storage to tape. Both the STKen and CDFen enstore robots were mounted and fcdfsam was granted permission to read/write to STKen and read from CDFen. The "enstore monitor" was run as a validation test after the encp product was installed to make sure the debug basic throughput to the enstore systems. A sam station was installed and the sam administration tools were deployed.

**fndaut**: a Sun was chosen for this machine since the SAM server software had been running for some time on this platform for DZero. The operating system was upgraded and 24x7 support arranged. A shift list with people from Fermilab Computing and CDF, both onsite and offsite was arranged. Sam database and CORBA servers were installed for development, integration and production. Specific tools installed were:

- CORBA name service (IOR)
- sam logger
- sam optimizer
- sam CORBA db server
- web servers
- cron jobs to update servers and web pages containing computations of statistics and monitoring tools.

**Client Test Stations**: two sam stations have been established to test and debug the deployed infrastructure: one at Fermilab (nglas09) and one in the UK (cdfa).

**Databases**: the D0 SAM schema was cut cleanly onto the CDF machines following the Fermilab computing division standards [6], first cutting the schema into development, then integration and finally production. Data required by the SAM middleware were loaded during the database cut. Other data were added with the "Predator" program described in the section below.

**Feature adjustments**: being able to access files via the DCache layer was a principal feature required by the CDF data handling group. SAM implements a plug in mechanism to accommodate a large variety of file transfer protocols. Before the phase of deployment, we developed the plug in that would enable access to DCache via its weakly authenticated ftp door. This authentication mechanism was acceptable for a read-only type of access. This development work helped in addressing the second of our goals: "Enabling access to datasets of interest". The same plug in has then been further developed to include access to the gridftp door directly or via SRM.

**Administration**: With a large number of people working across two continents communication was essential. This was maintained by using Polycom Viavideo PC-based video systems. Fermilab and UK institutes had these systems and met daily for 15 minutes to examine the task list and schedule one-on-one discussions amongst experts. This provided invaluable in ensuring the project kept pace.

## 5. MAPPING THE CDF DFC TO SAM

The CDF DFC was transported into SAM using the "Predator" program [7]. A first version of Predator mined data from the DFC and wrote python scripts that could be processed by the SAM command line API for storage of files. This was found to be cumbersome and much of the metadata that could be provided by CDF was not available at the command line. Therefore direct java-based access to the SAM schema was developed. This required development of a table-by-table and column-by-column mapping of the DFC to SAM schema. It was found that there were three categories of information that needed to be provided:

1. Information available from the DFC in the filename or from other parts of the CDF database that could be put into SAM as database entries.
2. Information not tracked in DFC but required by SAM: we decided to enter something logical.

3. Information in DFC not accommodated in SAM. We decided to translate this either by using the SAM parameters or file inheritance OR by using a "hook" to the CDF database with foreign keys.

## 6. THE INTERFACE TO THE ANALYSIS PROGRAMS

In order for CDF analysis programs operating in the AC++ framework [8] to access files from SAM an adaptor was written: the CDF Project Protocol Converter (CPPC). Since the CDF software does not use exception handling and the D0 framework does, it was not possible to link the CORBA interface software directly into the CDF software. It was decided that the AC++ module would fork a protocol adaptor and communicate with it using a finite state machine. Messages included: configure, getfile, and error. System parameters were communicated from the AC++ framework to the CPPC via pipelines. The main output from the SAM system was a fully qualified filename. The CDF FileInput C++ based class required only an "append" of the filename to the list to be processed and hence adaptation was quite easy. Further modifications to the adaptor were made to allow multiple processes to be started on parallel machines so that files could be delivered equally to all nodes. The user interface was quite easy: the user needed only supply a dataset and set an environment variable with a unique name for the project.

## 7. EXPERIENCE AND LESSONS LEARNED

A major issue in deploying SAM for a second experiment was that of controlling the versions of the various packages and ensuring that they would function together. This was exacerbated by the rapid growth in the number of sites concurrently using SAM. The FNAL UPS/UPD package distribution system allows for automatic installation of packages upon which other depend. Conflicting dependencies made it difficult to establish a compatibility matrix for the various SAM software components.